Correlative Study of Enhanced Excitonic Emission in ZnO Coated with Al Nanoparticles using Electron and Laser Excitation.

*Saskia Fiedler\*[1,2], Laurent O. Lee Cheong Lem[1,3], Cuong Ton-That[1], Markus Schleuning[4,5], Axel Hoffmann[4], Matthew R. Phillips\*[1]*

Dr. Saskia Fiedler, Dr. Laurent O. Lee Cheong Lem, Assoc. Prof. Cuong Ton-That, Prof. Matthew R. Phillips
1. School of Mathematical and Physical Sciences, University of Technology Sydney, 15 Broadway, Ultimo NSW 2007, Australia

Dr. Saskia Fiedler
2. Centre for Nano Optics, University of Southern Denmark, Campusvej 55, 5230 Odense M, Denmark

Dr. Laurent O. Lee Cheong Lem
3. Australian National Fabrication Facility, Australian National University, Canberra ACT 2601, Australia

Markus Schleuning, Prof. Axel Hoffmann
4. Technische Universität Berlin, Fakultät II, Institut für Festkörperphysik, Sekretariat EW 5-4, Hardenbergstr. 36, 10623 Berlin, Germany

E-mail: safi@mci.sdu.dk and matthew.phillips@uts.edu.au

Metal nanoparticle (NP) surface coatings are known to significantly enhance the ultra-violet luminescence intensity of ZnO. Although there is general agreement that resonantly excited Localized Surface Plasmons (LSPs) in metal NPs can directly couple to excitons in the semiconductor increasing their spontaneous emission rate, the exact mechanisms involved in this phenomenon are currently not fully understood. In this work, LSP-exciton coupling in a ZnO single crystal and ZnO nanorods coated with a 2 nm Al layer has been investigated using correlative photoluminescence and depth-resolved cathodoluminescence and time-resolved photoluminescence spectroscopy. Temperature-resolved cathodoluminescence and photoluminescence measurements from 10 K to 250 K show enhancement factors up to 12 times of the free exciton (FX) emission at 80 K. The FX couple more efficiently to the LSPs in Al compared to the localized donor-bound excitons. Furthermore, a strong polarization dependence of the LSPs with respect to the FX was observed with higher enhanced FX transitions polarized in the same direction as the electric field of the incident excitation. These results indicate that selective enhancement of the ultra-violet excitonic PL in ZnO can be achieved by careful alignment of the crystallographic axes of the ZnO relative to the electric vector of the excitation source.

**1. Introduction**

Localized surface plasmon (LSP) – exciton coupling has been widely reported to significantly enhance the luminescence of ZnO and other semiconductors by using a metal nanoparticle (NP) surface coating. [1–8] LSPs greatly enhance the local electric field at the surface of the metal of NPs via the collective oscillation of free electrons in the presence of an external electromagnetic field. Significantly, when the plasma resonance energy of the metal NPs matches the excitonic energy of the semiconductor, LSPs can efficiently couple to excitons in semiconductors. This LSP resonance energy is dependent on the type of metal employed, the dimension and the shape of the NPs as well as the permittivity of the surrounding dielectric. Silver (Ag) and gold (Au) are the most commonly used metals for plasmonic applications due to their low optical losses in the visible and near infrared spectral regions and their relative stability in air. However, Au is widely considered to be unsuitable in the UV as it has high losses for energies above 2.3 eV due to its interband transitions. Although, Ag seems eminently suitable for UV plasmonic applications as its interband transitions occur at higher optical energies, this is not the case as it is less stable in air than Au. Specifically, Ag readily forms oxide and sulphur layers when exposed to air making the fabrication of stable plasmonic Ag NPs difficult. [9,10] Conversely, Aluminum (Al) affords more robust NPs due to strong self-limiting oxidation effects even in very small particles. [11] Moreover, Al NPs have the advantage of a widely tunable plasmon resonance energy from the deep UV to the visible spectral range, while being cheaper and more abundant than both Au and Ag. [12–15] These advantages are clearly highly desirable for plasmonic applications in the blue-UV spectral range, in particular, the enhancement of excitonic luminescence in ZnO.

ZnO is a wide-bandgap semiconductor (Eg = 3.376 eV at 4 K) with a large exciton binding energy of 60 meV. [16–18] A typical ZnO luminescence spectrum consists of a sharp UV emission at around 3.37 eV and a broad visible defect-related emission positioned between 2.3 and 1.7 eV. [18] The UV emission at 4 K originates from excitonic recombination involving three free excitons (FX) ($E_A$ = 3.3781 eV; $E_B$ = 3.3856 eV; $E_C$ = 3.4262 eV) due to split valance band of ZnO as well as a number of donor-bound excitons (DBX), arising from impurities, such as H, Al, Ga and In. [19–21] The polarization of the FX-A and FX-B in wurtzite ZnO is different to FX-C: The FX-A and FX-B are perpendicularly polarized to the *c*-axis, while FX-C is parallel polarized. [22–24] As the temperature is increased from 80 K to 300 K, the DBXs thermally ionize and quench while the FX and its LO phonon replicas with a 72 meV spacing red shift, become wider and overlap to produce a broad near band edge emission centered at 3.30 eV.

The ZnO UV emission is greatly enhanced by the addition of a thin Al NP surface coating as the excitons in ZnO can directly couple to the LSPs in the Al NP. In this process, the presence of an additional, non-radiative, faster relaxation LSP channel results in an increased spontaneous emission rate and consequently, an enhanced UV excitonic luminescence intensity. This LSP-exciton coupling mechanism in an Al-coated ZnO nanorod (NR) is illustrated in **figure 1**. Although the LSP-exciton coupling between ZnO and Al interpretation for the increased emission seems plausible and is widely accepted, there is an inexplicably large range of enhancement factors reported in the literature, extending over orders of magnitude. [1,3,25–30] To address this issue, an extensive investigation of the coupling mechanism in Al NP coated ZnO has been conducted.

In this work, two types of Al-coated ZnO samples, *a*-plane ZnO single crystals and a vapor–solid (VS)-grown ZnO NR ensemble, were systematically studied using photoluminescence (PL) spectroscopy and depth-resolved cathodoluminescence (CL) spectroscopy at temperatures ranging from 10 K to 250 K. An up to 12 times enhanced UV PL emission of Al-coated ZnO was typically observed compared with their uncoated counterparts. In comparison, an 8 times enhancement was measured with a CL excitation of 3 kV. The Al LSPs were found to couple more strongly to the mobile FX in ZnO compared with localized DBX with the greatest enhancement observed with exciton excitation close to the surface.

Additionally, the LSP-exciton coupling enhancement strength of the FX-A/FX-B and FX-C emissions was found to be a dependent on the crystallographic orientation of the ZnO sample relative to the Al NP LSP dipole direction, which is mediated by the electric field alignment of the excitation source. Differences in the measured excitonic UV emission enhancement between PL and CL excitation mechanisms are attributed to variation in the FX-LSP coupling strength caused by different: (i) FX generation depths relative to the NP-ZnO surface and (ii) orientation of the LSP dipole relative to the surface plane.

**2. Results and Discussion**

Depth-resolved CL spectra of the uncoated and Al-coated ZnO at a temperature of 10 K and 80 K are shown in **figure 2**. At both temperatures, an enhanced UV emission from the Al-coated side of the sample is observed compared to the adjacent uncoated ZnO control region. The UV enhancement factor is obtained by dividing the integrated UV emission from 3.00 eV to 3.45 eV from the Al-coated ZnO by that of the uncoated side of the sample. At a temperature of 10 K, the UV emission enhancement factor at 3 kV and 5 kV is found to be 1.8, which reduced to 1.5 at an acceleration voltage of 10 kV. A higher UV enhancement close to the ZnO-Al interface is found at 80 K with integrated UV enhancement factors of 8.3 at 3 kV, 3.3 at 5 kV and 2.2 at 10 kV. This increased near-surface enhancement effect at 80K is illustrated in **figure. 3**, which shows the CL enhancement factor as a function of emission energy at 3 kV, 5 kV and 10 kV. Here, the dashed vertical lines show the FX-related peaks in ZnO, while the dotted lines indicate the DBX-related emissions. [20,21,31] Figure 3 shows that the first three LO phonon replicas of the FX-C emission at 3.353 eV, 3.280 eV and 3.203 eV are enhanced by the presence of the Al film. The plot also reveals a strong enhancement at 3.358 eV which corresponds to the donor bound exciton, $I_6$ line.[31] This emission peak has been unequivocally assigned to Al donor impurities in ZnO, suggesting that some Al may have in-diffused into the near surface region of the crystal. The arrow label at 3.343 eV in Figure 3 indicates an emission which cannot be attributed to any of the reported luminescence peaks in ZnO and $Al_2O_3$. As the enhancement level of this peak is observed to increase with decreasing kV, where carriers are injected closer to the surface, it is most likely due to an emission arising from interface related defects residing at the ZnO and Al boundary. Additionally, these results collectively confirm that the LSPs in Al couple more efficiently to the FX than to the DBX. Furthermore, the CL enhancement is temperature-dependent with the UV enhancement being greater at 80 K than at 10 K.

The temperature dependence of the Al induced UV enhancement of the PL from 10 K to 250 K is displayed in **figure 4**. Shown at the bottom of the plot are the measured PL spectra of the uncoated and the Al-coated samples at 10 K used to determine the energy dependent PL enhancement in the top frame. The dotted vertical lines specify the DBX-related emissions, while the dashed lines indicate the FX peaks. The enhanced emission peak at 3.343 eV indicated by the arrow label in figure 4 is consistent with the peak at the same energy observed in the CL measurements (figure 3) originating from radiative defects located at the ZnO-Al interface. The enhancement is found to maximize at three energy positions that are in agreement with the positions of the FX-C-LO1, FX-C-LO2 and FX-C-LO3 excitonic emissions. Moreover, these three maxima red shift in energy as the temperature increases from 10 K – 250 K which is consistent with the widening of band gap due to increasing electron-phonon scattering effect. The temperature-resolved data reveal that the maximum enhancement factor of ~ 12 was achieved at 80 K, following the FX-C emission intensity dependence on temperature. As the temperature rises from 10 K to 80 K, the competitive DBX recombination channels thermalize and quench, increasing the FX-C intensity, which then falls approaching 250 K as higher order exciton LO phonon replicas become dominant. This observation supports the assignment of the Al thin film induced emission enhancement to the formation of an additional relaxation channel due to LSP coupling to the FX-C.

Correlative temperature-resolved CL at 5 kV and PL spectroscopy measurements at 10 K and 80 K are shown in **figure 5(a)** and **5(b)**, respectively. At 10 K, the UV enhancement factor in CL and PL is similarly low at ~ 3.5. While the low energy side of the spectrum of the enhancement factor appears very similar for both types of excitation, the higher energy side above 3.334 eV shows a more pronounced UV enhancement in CL than in PL. This can be attributed to a higher contribution of the FX-A and FX-B LSP-exciton coupling when excited by the electron beam rather than by the laser. With both CL and PL excitation, greater Al LSP coupling occurs with FX compared with DBX in ZnO. This is because the FX have a larger spatial extent providing a larger wavefunction overlap than with the DBX, which are spatially localized at the donor. A similar argument is also used to explain why FX phonon coupling is generally much stronger the DBX-phonon coupling.[32,33] Greater FX over DBX coupling can also be due to the FX having a high mobility enabling FX diffusion towards the ZnO and Al surface coating interface, where the LSP-FX separation is small, and coupling is strongly enhanced.

Due to the higher FX emission intensity at 80 K, where the DBX is fully thermally dissociated, a greater coupling strength in PL and CL is observed, as shown in figure 5b. Here, the PL enhancement has its maximum of approximately 12, while the maximum CL enhancement is 4. The 3-fold difference in CL and PL enhancement can be explained by the difference in type of excitation: electron beam or laser light. While the excitation profile of the laser is highest at the surface following the Beer-Lambert law profile with depth where approximately 8% of the total optical absorption occurs in the first 5 nm of the sample. In contrast, Monte Carlo CASINO modelling shows that the maximum excitation in CL is deeper in the bulk of the sample at around 1/3 of the electron range. In CL at 5 kV, only about 2% of the energy is lost within the first 5 nm of the sample indicating that the generation of the FX close to the Al NP surface coating is considerably greater when excited by the laser compared with the electron beam. Therefore, the probability of LSPs in Al coupling to FX in ZnO is higher for laser than with CL.

Furthermore, the relative enhancement factors of FX-A, FX-B, and FX-C in Al-coated ZnO are different for laser and electron beam excitation. It is well-known that the polarization of the FX in ZnO are different for the FX-A and FX-B, which are perpendicularly polarized to the *c*-axis of ZnO, while the FX-C is parallel polarized. [22–24] This results in different LSP-exciton coupling strengths for the two different types of excitation. The polarization of the laser excitation is always perpendicular to the k-vector of the incident light; consequently, in our PL setup geometry the laser electric vector is always parallel to the sample surface. Accordingly, the surface parallel polarized LSPs in Al, excited by the laser can couple more efficiently to the equally polarized FX-C while the perpendicular polarized FX-A and -B cannot couple as strongly. Therefore, a higher enhancement of FX-C than that of FX-A and -B can be achieved with laser light excitation. Conversely, for electron excitation with the beam normal to the surface, the incident electron and its image charge within the NP is dipole-like, producing an electric field perpendicular to the surface efficiently which excites Al NP LSPs in the same direction. [34–36] Consequently, LSPs in the Al NPs are excited parallel to the surface under laser excitation and perpendicular to the surface with the electron beam. The LSP dipole orientation difference between PL and CL excitation results in higher enhancement of the FXs that are polarized in the same plane as the incident electric field of the excitation source.

A direct comparison of the CL and PL enhancement factors as a function of energy at 10 K and 80 K is displayed in figure 5a and 5b, respectively. At 10 K, the PL enhancement factor of the FX-C is approximately 3.5 while the FX-A and FX-B are close to 1. At 80 K, the PL shows an even stronger polarization effect with the FX-C being enhanced 12 times and the FX-A and FX-B showing a much lower enhancement of approximately 3 and 5, respectively. In contrast, the CL enhancement factors are found to be generally lower. Here, the Al-LSPs excited by the electron beam are polarized normal to the sample surface which is parallel to the polarization of the FX-A and FX-B but perpendicular to that of the FX-C. Thus, the combined CL enhancement of FX-A and FX-B is expected to be higher than that of FX-C. It is noteworthy that the LSP-coupling to FX-A and FX-B has a similar probability as these two relaxation channels are competing. This results in a similar polarization effect of the LSP-exciton coupling in CL as well as in PL, however, the overall enhancement is more pronounced in PL due to the near-surface excitation of the laser.

To further investigate this polarization-dependent enhancement, PL and CL measurements were taken on *c*-axis-oriented ZnO nanorods. **Figure 6a** shows the SE image of well-aligned ZnO nanorods grown by the vapor-solid (VS) method on a Si substrate. As shown in the XRD profile in **figure SI 1**, the nanowires with an average diameter of approximately 100 nm are grown predominantly along the (002) axis, perpendicular to the substrate. In this case, the FX-A and FX-B are parallel polarized to the sample surface while the FX-C is perpendicular polarized, resulting in the opposite polarization geometry to the *a*-plane ZnO single crystal discussed above. The LSP-exciton coupling of the ZnO NRs coated with a 2 nm thin Al film (cf. above) were studied at a temperature of 80 K to maximize the number of FX by dissociating DBX and thereby producing a higher enhancement factor. The near-surface excitation by laser light was chosen as it was found that electron beam excitation introduced charging effects in the ZnO NR sample. Furthermore, possible excitation of LSPs in Al NPs located at the side walls of the ZnO NRs by a sufficiently energetic electron beam can result in a 90-degree tilted polarization state of the Al-LSPs, and therefore, lead to an overall mixed polarization of LSP-exciton coupling.

**Figure 6b** shows the PL spectra of the uncoated and the Al-coated ZnO NRs at a temperature of 80 K, as well as the resulting enhancement factor as a function of energy. The FX-A and FX-B related emissions are highly enhanced as expected since the FX-A and FX-B are polarized in the same plane as the sample parallel to the electric vector of the incident laser light. The FX-C enhancement is less pronounced as its polarization is normal to the *c*-plane. These polarization studies confirm that the PL coupling strength of the Al NP LSPs to the three FX transitions in ZnO and their emission enhancement is determined by the crystallographic orientation sample, which fixes the polarization direction of the three FX with respect to the incident laser polarization. It is also noteworthy, that the greatest PL enhancement in the VS ZnO NRs is found at the Al DBX $I_6$ line, suggesting that Al may have diffused into the sample.

To confirm LSP-exciton coupling, TR-PL was performed on the planar a-plane ZnO single crystals. It is widely accepted that LSP-exciton coupling increases the spontaneous emission rate of the metal nanoparticle-coated ZnO through the creation of an additional, faster relaxation channel via the LSPs, leading to a reduction of the carrier lifetime in these samples. [1,37,38]

**Figure 7** shows the time-resolved PL results for both the uncoated and Al-coated *a*-plane ZnO crystal collected at 10 K. Deconvolution of the laser response with the signal was performed to obtain the fit of the relaxation curve to determine the lifetime of the emission. The uncoated side of the sample, illustrated in figure 7a, exhibits an excitonic lifetime of $\tau_{ZnO} = (161 \pm 4)$ ps, while figure 7b displays the TR-PL results of the Al-coated ZnO with a lifetime of $\tau_{ZnO+Al} \leq 53$ ps. The faster lifetime from the Al-coated sample is consistent with an increase in the spontaneous emission rate through the creation of an additional, faster relaxation channel via the LSP-exciton coupling, leading to a reduction of the carrier lifetime. The lifetime of the Al-coated ZnO is very close to the laser response, which does not allow the decay curve to be confidently fitted. However, a maximum lifetime of 53 ps can be used to calculate the Purcell enhancement factor, $F_P$, which quantifies the coupling strength between the LSPs in the Al nanoparticles and excitons in ZnO.

$$F_P = \tau_{ZnO}/\tau_{ZnO+Al} \geq 3.0 \qquad (1)$$

Due to the limited temporal resolution of the TR-PL constrained by the laser response, a lower limit of the Purcell enhancement factor was determined to be $F_P \geq 3.0$. A direct comparison of this $F_P$ value to the maximum PL enhancement of 12 times at a temperature of 80 K is not trivial due to various reasons. First, the collection of the TR-PL data was only possible at a temperature of 10 K where the PL spectrum of ZnO is dominated by the DBX which couple less effectively to the LSPs in Al than the FX. The corresponding experimentally derived PL enhancement factor is, therefore, as low as 4 (cf. figure 4), indicating that the calculated $F_P$ is on the same order. Second, the limited temporal resolution of the TR-PL setup only allows the determination of a lower limit of the Purcell enhancement factor not allowing for a quantitative comparison with the experiment. However, the more than three times reduced lifetime in Al-coated ZnO single crystals clearly confirms an increased spontaneous emission rate via LSP-exciton coupling.

## 3. Conclusion

In conclusion, LSP-exciton coupling in two types of Al-coated ZnO samples - *a*-plane ZnO single crystals and VS-grown ZnO nanorods - was studied using CL and PL spectroscopy. The results reveal that the observed enhancement is strongly dependent on the type of excitation, electron or photon, and the sample temperature. The greatest UV enhancement of approximately 12 times was observed at a temperature of 80 K using laser excitation. Due to the higher mobility and the larger spatial extent of the FX in ZnO compared with the DBX, a stronger coupling of the LSPs in the Al to the FX in ZnO was found. At elevated temperatures, the LSP-exciton coupling is more effective, where the spectrum is dominated by FX emissions as the DBX are mostly thermally dissociated. CL spectroscopy confirmed that the maximum enhancement was observed at the surface close to the ZnO-Al interface with a higher enhancement factor at 80 K than at low temperatures.

Furthermore, it was demonstrated that the ZnO crystal orientation with respect to the electric vector orientation of the incident laser light determines which of the three FX transitions are coupled more strongly to the LSPs. The strongest PL enhancement of the FX emission was observed when the FX polarization in the same plane as the electric vector of the incident laser excitation.

## 4. Experimental Section

*Fabrication of ZnO samples:*

In this work, hydrothermally grown, 0.5 mm thick, 10 x 10 mm *a*-plane ZnO single crystal polished plates were obtained from MTI Corp (USA).

Hexagonal ZnO NR ensembles with a mean diameter of ~ 100 nm and a length of ~ 1500 nm were grown on *a*-plane sapphire substrates using a vapor solid technique, which is described in detail elsewhere. [39]

Both the single crystal and ZnO NR samples were sputter-coated at room temperature with a 2 nm thin Al film. The resulting surface coating consists of metallic Al NPs embedded in an $Al_2O_3$ matrix as shown by ellipsometry and a reduced UV transmission (see **Figure SI 2**). Prior to deposition, the single crystal samples were subjected to a standard cleaning procedure consisting of a 20-minute sonication in acetone, then isopropyl alcohol and lastly deionized water. Only half of each sample was coated with Al to leave the other, uncoated side as a control reference to facilitate a direct before and after deposition comparison of the luminescence intensity.

*Optical characterization:*

CL and PL spectroscopy were performed in a FEI Quanta 200 SEM equipped with a Gatan CF302 continuous flow liquid helium cold stage, allowing for temperature-dependent measurements from 10 K to room temperature. Light injection optics allow sequential CL and PL measurements on spatially equivalent regions of the samples without breaking the SEM vacuum. PL was excited with the 325 nm line of a Melles Griot He-Cd laser with a power of 2.1 mW. CL and PL emitted from the sample were collected by a parabolic mirror and analyzed using either an Ocean Optics QE Pro spectrometer for full spectral range or a Hamamatsu S7011-1007 CCD for high spectral resolution spectroscopy. All spectra were corrected for the total response of the system. SEM acceleration voltages of 3 kV, 5 kV, and 10 kV were used for depth-resolved CL analysis and the SEM beam power was kept constant at 17.5 μW by varying the electron beam current. [40] The Monte Carlo CASINO simulation package was used to determine the in-depth spatial distribution of the injected electron hole pairs.[40] For SEM acceleration voltages of 3 kV, 5 kV and 10 kV, the simulations reveal CL generation depths of approximately 40 nm, 100 nm and 350 nm, respectively, corresponding to 70% of the total electron energy loss. At an acceleration voltage of 5 kV, the CL excitation depth is very similar to the excitation range of the 325 nm UV laser in ZnO of ~ 100 nm. However, the electron hole pair injection density with CL excitation is approximately three orders of magnitude larger than that of the laser at a power of 2.1 mW.

Time-resolved PL was performed at 10 K and the luminescence decay was recorded by a standard photon counting technique. In these measurements, the excitation wavelength of the pulsed fiber ($\lambda = 1031$ nm) laser was internally halved and subsequently frequency-doubled to $\lambda = 258$ nm with a pulse duration of 5.5 fs at repetition rate of 76 MHz.

**Acknowledgements**
The authors would like to thank Katie McBean, Geoff McCredie and Dr. Angus Gentle for the technical support at UTS, and Dr. Stefan Kalinowski for his assistance at TU Berlin. Furthermore, we would like to acknowledge financial support from the Australian Research Council (DP150103317).

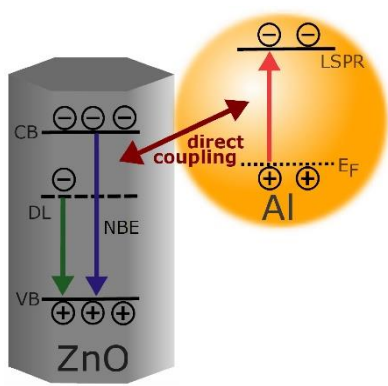

**Figure 1.** Illustration of the LSP-exciton coupling mechanism.

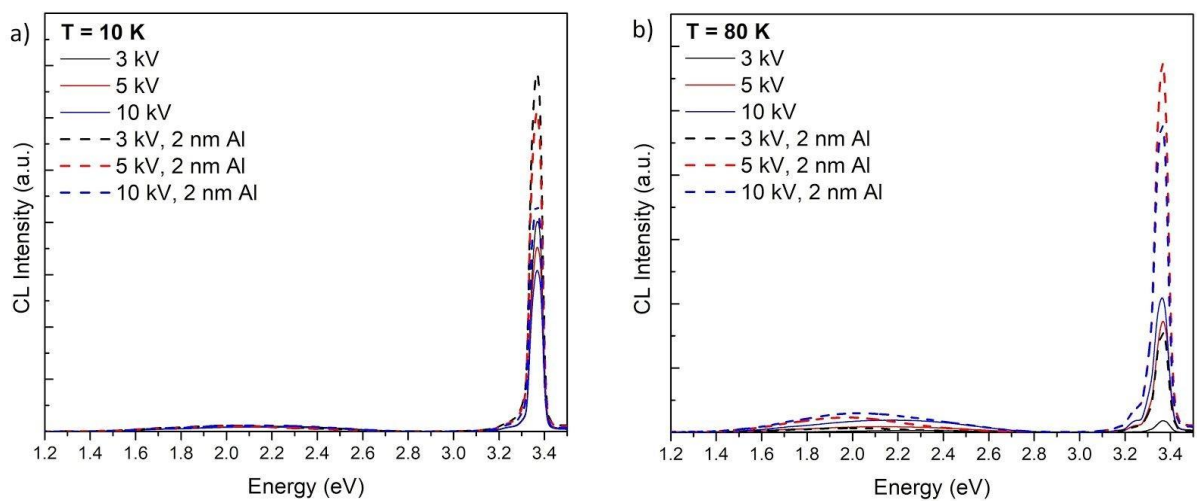

**Figure 2.** Depth-resolved CL of uncoated (solid lines) and Al-coated ZnO (dashed lines) at a temperature of (a) 10 K and (b) 80 K. $P$ = 17.5 μW with varying accelerating voltage of 3 kV, 5 kV and 10 kV, corresponding to approximate CL generation depths of 40 nm, 100 nm and 350 nm.

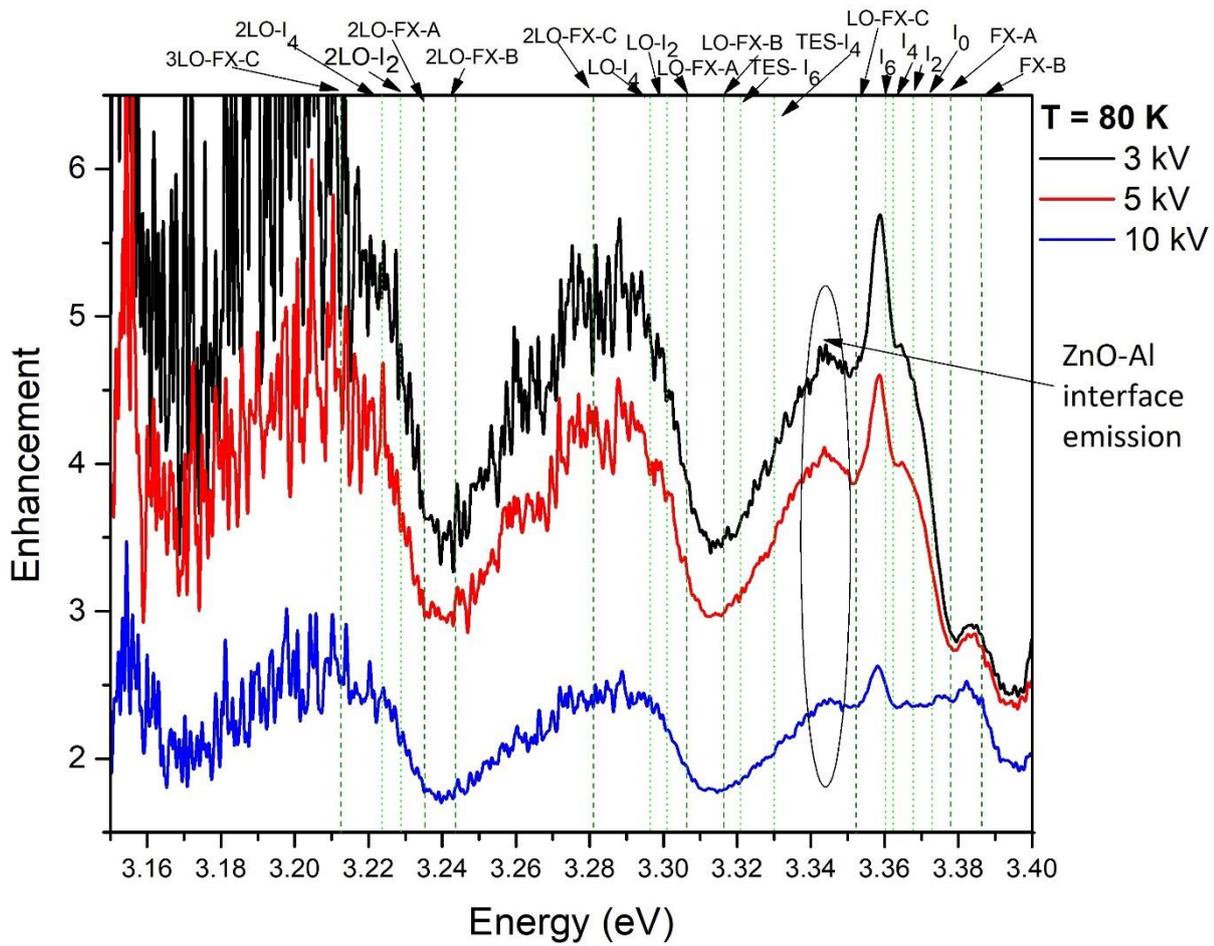

**Figure 3.** CL enhancement factor as a function of energy at acceleration voltages of 3 kV (black), 5 kV (red) and 10 kV (blue), resulting in CL generation depths of approximately 40 nm, 100 nm, and 350 nm, respectively. Dashed green lines illustrate the DBX-related transitions in ZnO, while the dotted olive lines represent the FX-related emissions. $P$ = 17.5 μW, $T$ = 80 K.

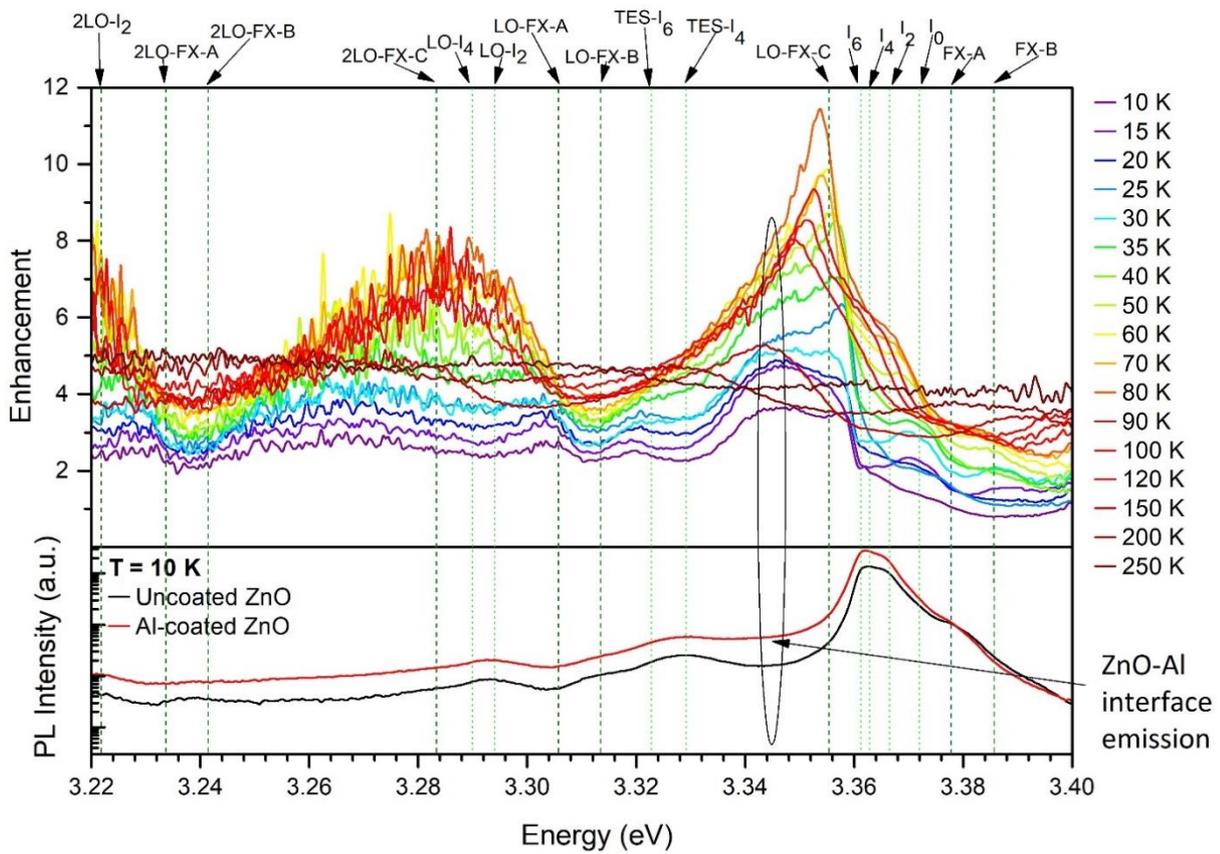

**Figure 4.** Bottom: PL spectra of the uncoated (black) and Al-coated (red) ZnO at 10 K. Top: Temperature-resolved PL enhancement factor as a function of energy. Dashed green lines illustrate the DBX-related transitions in ZnO, while the dotted olive lines represent the FX-related emissions. $P$ = 2.1 mW.

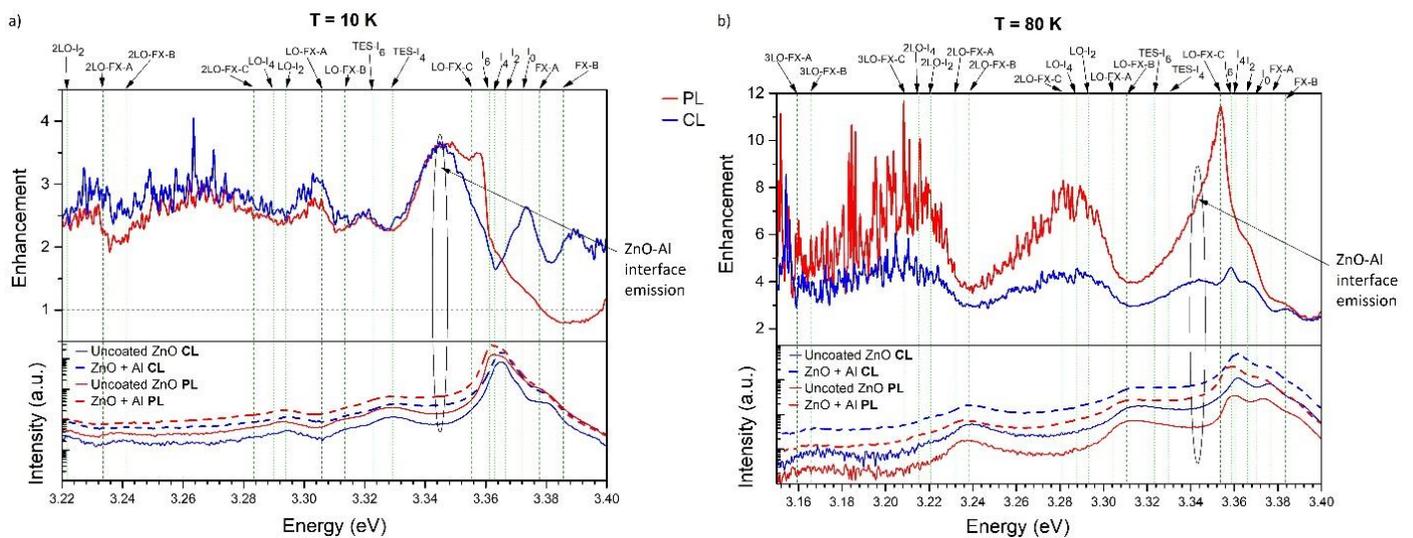

**Figure 5.** (a) Bottom: PL (blue) and CL (red) spectra at 5 kV of the uncoated (solid lines) and Al-coated (dashed lines) ZnO at 10 K. Top: Corresponding PL and CL enhancement factor as a function of energy. Dashed green lines illustrate the DBX-related transitions in ZnO, while the dotted olive lines represent the FX-related emissions.

(b) PL and 5 kV CL spectra and enhancement factors at a temperature of 80 K.

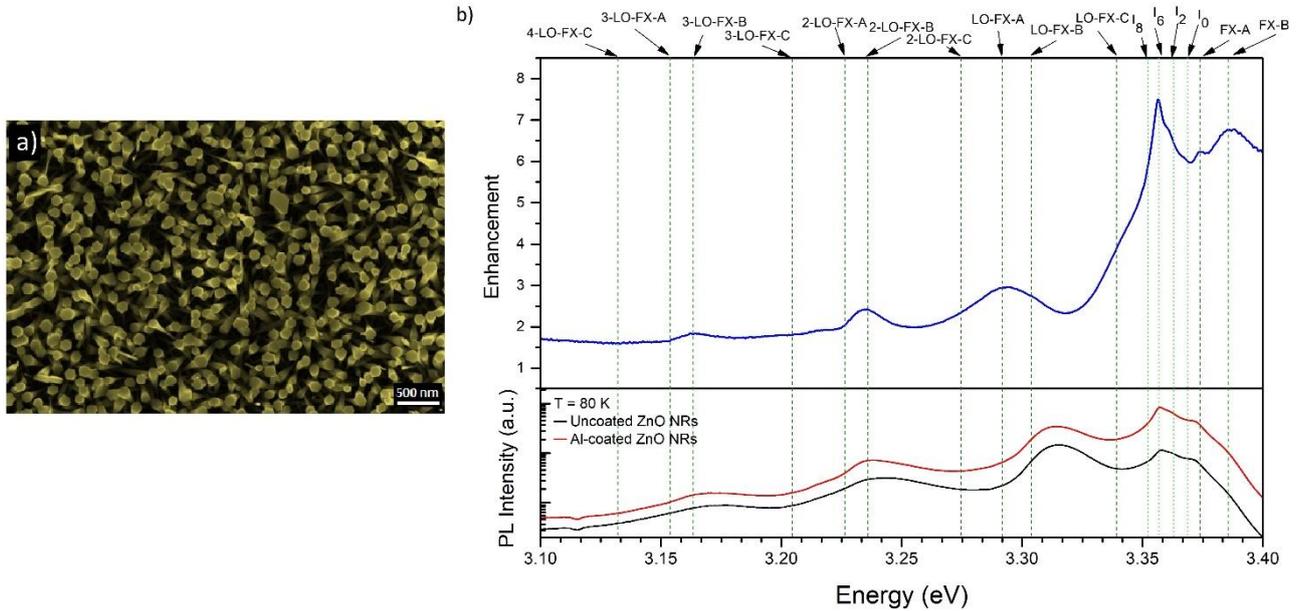

**Figure 6.** (a) SE image of the uncoated VS-grown ZnO nanorods with an average diameter of 100 nm.

(b) Bottom: PL spectra of the uncoated (black) and Al-coated (red) VS-grown ZnO nanorods at 80 K. Top: Corresponding PL enhancement factor as a function of energy. Dashed green lines illustrate the DBX-related transitions in ZnO, while the dotted olive lines represent the FX-related emissions.

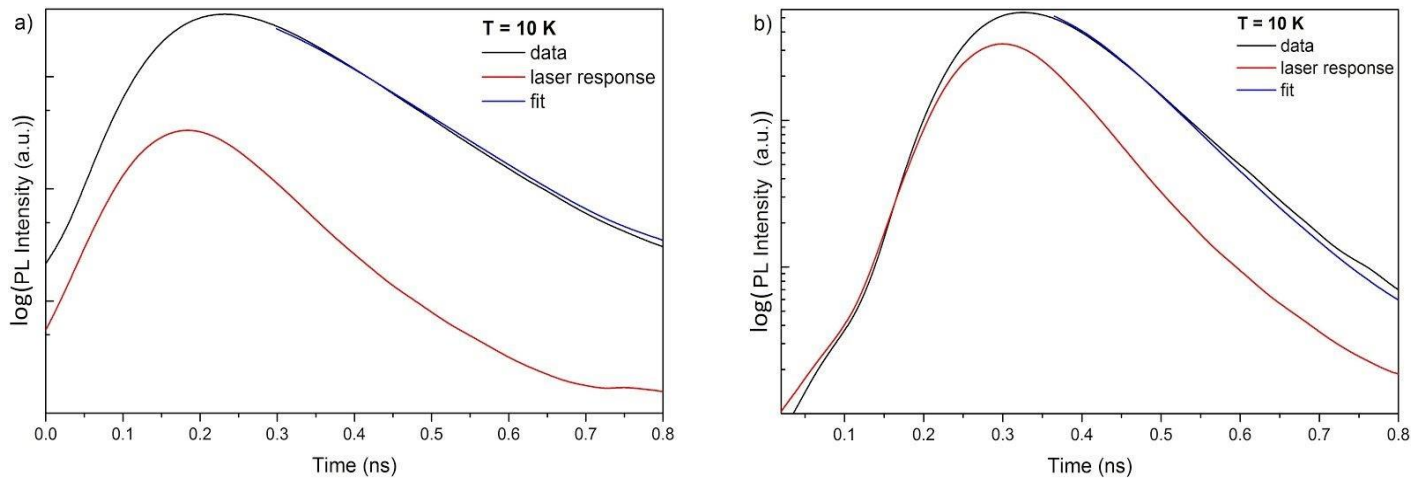

**Figure 7.** (a) Time-resolved PL of the uncoated ZnO with a lifetime of $\tau_{ZnO} = (161 \pm 4)$ ps and (b) of the Al-coated ZnO, showing a reduced lifetime of $\tau_{ZnO+Al} \leq 53$ ps.

**Supporting Information**

Supporting Information is available from the Wiley Online Library or from the author.

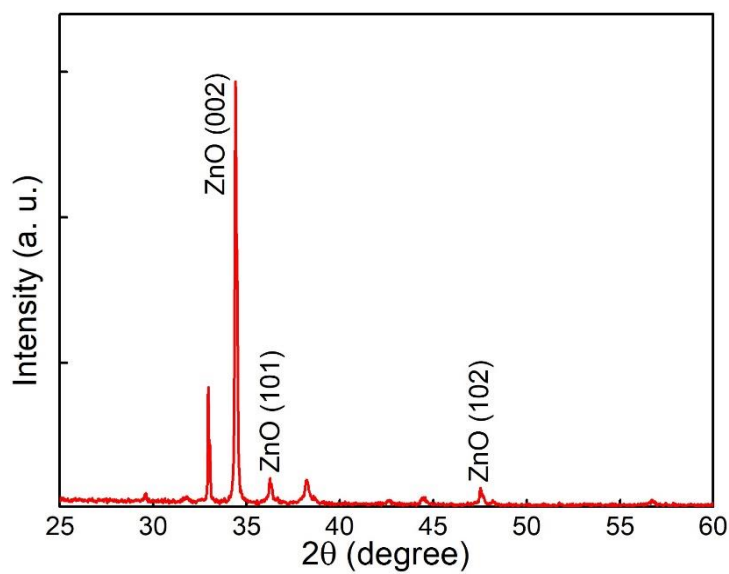

**Figure SI 1.** XRD spectrum of VS-grown ZnO nanorods on a Si substrate with ZnO (002) peak being most dominant indicating growth along *c*-axis.

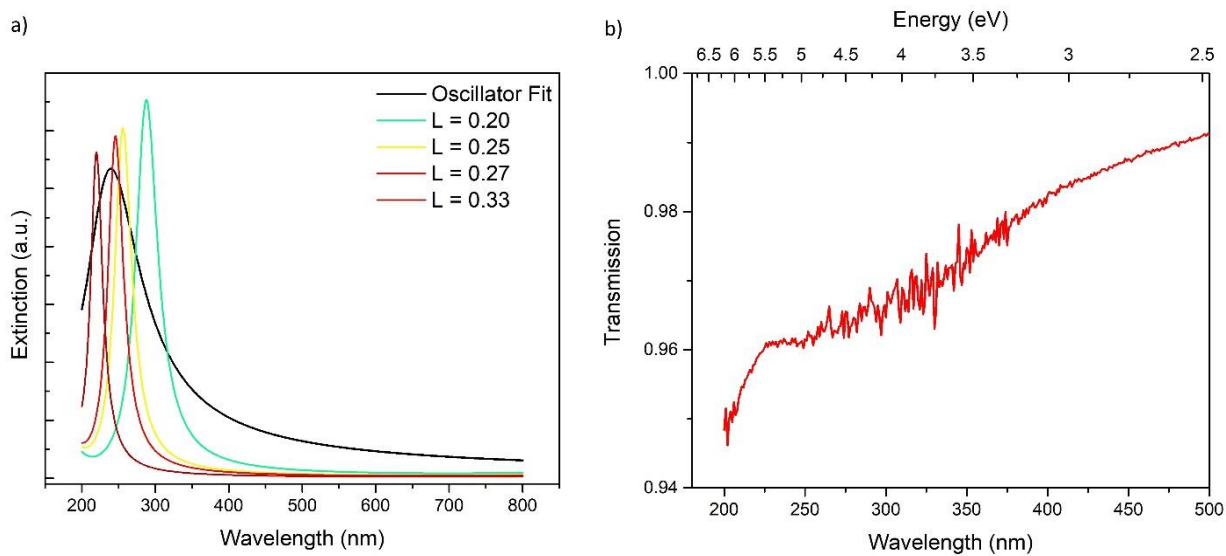

**Figure SI 2**. (a) Extinction spectrum of a 2nm thin Al film on a Si substrate fitted with an oscillator model, obtained by ellipsometry (black). Modelled extinction spectra of differently shaped Al nanoparticles in an $Al_2O_3$ matrix (colored lines), where L = 0.33 corresponds to a spherical nanoparticle and L = 0.20 a nanorod. The other L-values correspond to a more elliptical shape, being between those two L-values, indicating that the sample Al hot contains a large distribution of differently shaped Al nanoparticles. (b)Transmission spectrum of Al surface coating

on UV-quartz divided by the reference spectrum of uncoated UV-quartz, showing LSPR absorption in the UV region of the spectrum.